\documentclass[useAMS,usenatbib]{mn2e}
\usepackage{lscape}
\usepackage{url} 
\usepackage[dvips]{graphicx}
\usepackage{array}
\usepackage{aas_macros}
\usepackage{upgreek}
\usepackage[T1]{fontenc}
\usepackage{times}
\usepackage{amsmath}
\usepackage{aecompl} 
\usepackage{multicol}
\usepackage{multirow}
\usepackage{mathtools}
\usepackage{nicefrac} 
\usepackage{ifpdf}
\usepackage{xcolor}
\ifpdf
  \usepackage{epstopdf}
\fi

\def\gtrsim{\mathrel{\hbox{\rlap{\hbox{\lower3pt\hbox{$\sim$}}}\hbox{\raise2pt\hbox{$>$}}}}}

\newcolumntype{R}{>{\raggedleft\arraybackslash}p{3cm}}
\newcolumntype{x}[1]{>{\centering\let\newline\\\arraybackslash\hspace{0pt}}p{#1}}
\title[Shapiro delay in PSR~J1811$-$2405]
  {A Shapiro delay detection in the pulsar binary system PSR~J1811$-$2405}
\author[C. Ng et al.]
  {C.~Ng$^{1,2}$\thanks{E-mail: cherry.ng@dunlap.utoronto.ca}, L.~Guillemot$^{3,4}$, P.~C.~C.~Freire$^2$, M.~Kramer$^{2,5}$, D.~J.~Champion$^2$,
\newauthor
  I.~Cognard$^{3,4}$, 
  G.~Theureau$^{3,4,6}$,  E.~D.~Barr$^{2,7}$\\
  $^1$Dunlap Institute for Astronomy and Astrophysics, University of Toronto, 50 St. George Street, Toronto, ON M5S 3H4, Canada\\
  $^2$Max-Planck-Institut f\"{u}r Radioastronomie, 
      Auf dem H\"{u}gel 69, D-53121 Bonn, Germany \\
  $^3$Laboratoire de Physique et Chimie de l'Environnement et de l'Espace, 
      LPC2E, CNRS-Universit\'{e} d'Orl\'{e}ans, F-45071 Orl\'{e}ans, France \\
  $^4$Station de Radioastronomie de Nan\c{c}ay, 
      Observatoire de Paris, CNRS/INSU, F-18330 Nan\c{c}ay, France \\
  $^5$Jodrell Bank Centre for Astrophysics,
      University of Manchester, Alan Turing Building,
      Oxford Road, Manchester M13 9PL, United Kingdom \\
  $^6$LUTH, Observatoire de Paris, PSL Research University, CNRS, Universit\'{e} Paris Diderot, Sorbonne Paris Cit\'{e}, 92195, Meudon, France \\
  $^7$Centre for Astrophysics and Supercomputing, 
      Swinburne University of Technology, 
      Mail H30, PO Box 218, VIC 3122, Australia }
\date{Released 2020 Xxxxx XX}

\pagerange{\pageref{firstpage}--\pageref{lastpage}} \pubyear{2020}

\def\LaTeX{L\kern-.36em\raise.3ex\hbox{a}\kern-.15em
    T\kern-.1667em\lower.7ex\hbox{E}\kern-.125emX}

\begin{document}

\label{firstpage}

\maketitle

\begin{abstract}
This paper presents the first detection of Shapiro delay from the binary millisecond pulsar PSR~J1811$-$2405. 
We report a 11-$\sigma$ measurement of the orthometric amplitude, 
$h_{3}=6.8(6)\times10^{-7}$, 
and a 16-$\sigma$ measurement of the orthometric ratio,  $\varsigma=0.81(5)$. 
Given the relatively high orbital inclination,  $i=79(2)^{\circ}$, of this binary system, 
we obtain constraints on the companion mass of $m_{\rm{c}}=0.31\substack{+0.08 \\ -0.06}\,M_{\odot}$. 
The pulsar mass is currently less well constrained, with a value of 
$2.0\substack{+0.8 \\ -0.5}\,M_{\odot}$.
The companion mass and the orbital period are in agreement with the prediction made by
previous numerical calculations of the evolution of compact binary systems. From a study of the polarization, 
we find that the orbital inclination angle is $\sim$100$^\circ$ and that PSR~J1811$-$2405 is an orthogonal rotator.
In addition, the $\upmu$s-level timing precision together with its narrow profile make PSR~J1811$-$2405
 a good candidate for inclusion in the pulsar timing arrays being used to detect nHz Gravitational waves.  
\end{abstract}


\section{INTRODUCTION} \label{sec:intro}
PSR~J1811$-$2405 was discovered by the High Timing Resolution Universe (HTRU) Pulsar survey \citep{HTRU1} conducted with the 64-m Parkes radio telescope. This millisecond pulsar binary system was initially published by \citet{HTRU2}. PSR~J1811$-$2405 was reported to be a typical low-mass binary pulsar likely with a Helium white dwarf (He-WD) companion in an orbit of roughly 6.27$\,$d. No post-Keplerian (PK) parameters were detected at the time of the previous publication. 

Using the best available radio timing ephemeris, \citet{Ng2014} 
detected gamma-ray pulsations from this pulsar in 
data from the \textit{Fermi Gamma-ray Space Telescope} \citep[Fermi;][]{FermiLAT}. 
In order to keep the radio ephemeris up to date and allow for accurate folding of
all the gamma-ray data, this pulsar is observed regularly with the
Nan\c{c}ay Radio Telescope.
Thanks to this as 
well as a dedicated timing campaign with the Effelsberg Radio Telescope, 
we have timing data spanning a total of 6.9\,yr.
As a result of this timing project, we have detected the Shapiro delay for this system,
and showed that it has a fairly edge-on orbital inclination. 

First proposed by \citet{Shapiro1964}, the Shapiro delay is the retardation 
in the arrival times of a pulsar's pulses as they propagate through the curved 
space-time in the close vicinity of the binary companion. This effect is most 
readily observed in binary systems viewed nearly edge-on, and in some cases 
allows for precise measurements of the pulsar and companion masses, 
e.g. PSR~J1614$-$2230 \citep{Demorest2010}.
This and PSRs~J0348+0432 and J0740+6620, with implied masses of
$1.908\, \pm \, 0.016$, $2.01 \, \pm \, 0.04$ and $2.14\, \pm \, 0.10 \, M_{\odot}$
respectively
\citep{NANOGrav11,Antoniadis2013,Cromartie2019}, have effectively 
ruled out a number of equations of state for matter at densities above those of the atomic nucleus
(EOS), including almost all currently proposed  hyperon or boson condensate EOSs. This
represents a very important constraint on the unknown state of matter at those
densities \citep{Ozel2016}.
Nonetheless, precise neutron star mass measurements in general are hard to achieve 
because suitable systems are rare: Thus far, only 43 such measurements have been made
\footnote{http://www3.mpifr-bonn.mpg.de/staff/pfreire/NS\_masses.html}.

It is thus of great interest to obtain more neutron star mass measurements, 
in order to probe a wider parameter space of the EOS and to better understand 
the true distribution of neutron star masses.

This paper is organized as follows: in Section~\ref{sec:obs} we describe 
the radio timing observations conducted for PSR~J1811$-$2405, including 
the instrumental set-up and the derivation of times-of-arrival. 
In Section~\ref{sec:results} we present our updated timing solution and 
discuss results from the polarization study as well as from the Bayesian $\chi^{2}$ 
analysis of the masses of the pulsar and its companion. 
We summarise our work in Section~\ref{sec:conclusion}.

\section{RADIO TIMING OBSERVATIONS} \label{sec:obs}

\subsection{Observational set-up}
Observations at the Nan\c{c}ay Radio Telescope (NRT) began in 2012 as part of 
their regular gamma-ray pulsar follow-up timing campaign. 
The Nan\c{c}ay Ultimate Pulsar Processing Instrument \citep[NUPPI;][]{NUPPI} 
was employed to take coherently dedispersed timing data.
The NUPPI data set has high timing precision and is the longest span data set we have. 
\begin{table}
\setlength{\tabcolsep}{0.13cm}
 \caption{Specifications of the observing systems employed for the 
timing observations in this work. 
$f_{\rm{c}}$ represents the associated receiver central frequency and $B$ is the backend usable bandwidth both in MHz. 
The number of TOAs generated from each data set is listed in the last column.}
 \begin{minipage}{9cm}
\begin{tabular}{llll}
  \hline
  Backend       &  Receiver $f_{\rm{c}}$ (MHz) & $B$ (MHz)& No. TOAs  \\
\hline
  Effelsberg PSRIX  & 1347           & 200        &  90 \\ 
  Nan\c{c}ay NUPPI & 1484          & 512         & 2368 \\ 
  Nan\c{c}ay NUPPI  & $\sim$2200 & 512 & 74 \\
\hline \label{tab:specs}
 \end{tabular}
 \end{minipage}
\end{table}

Despite its high instrumental sensitivity, the meridian nature of the NRT meant 
that PSR~J1811$-$2405 could not be tracked for more than an hour. 
Hence we conducted an intense timing campaign with daily observations during the 
6.27$\,$d orbit with the Effelsberg 100-m Radio Telescope in November 2015.
This campaign was designed to track the pulsar as long as possible on 
November 19 during superior conjunction. 
Shorter tracks were taken either side of the superior conjunction 
observation to ensure good orbital coverage and 
 to increase the confidence level of the Shapiro delay measurement. 
The Effelsberg observations were taken at the 1.4-GHz band using the central feed of the 
7-beam receiver (P217mm).
We used the PSRIX backend \citep{Lazarus2016}
which is based on the ROACH board and collects data in coherent dedispersion folding mode.  
One 1.6-hr scan was carried out at a wavelength of 6\,cm using the S60mm receiver  
although no pulsar signal was detected, setting an upper limit of flux density 
at $S_{\rm{4.9GHz}}<$0.02\,mJy. 
Refer to Table~\ref{tab:specs} for the specifications of all the receivers 
employed in this timing analysis.

\subsection{Derivation of times-of-arrival and timing solution} \label{sec:timing}
The \textsc{psrchive} data analysis package \citep{Hotan2004} was used for the 
radio timing data reduction. 
Each observation was corrected for dispersion and folded at the predicted topocentric pulse period. 
We co-added high signal-to-noise observations and created noise-free reference template for each observing frequency using wavelet smoothing as implemented in the {\tt psrchive} program of  {\tt psrsmooth} \citep{Demorest2015}.
We convolved the template with each individual profile to produce a time-of-arrival 
(TOA) \citep{Taylor1992}. Multiple TOAs per observation were generated when 
possible, by downsampling the bandwidth to four frequencies and summing over every 10\,min. 
This is appropriate for the accurate measurement of orbital effects such as Shapiro delay 
for a 6-d binary like PSR~J1811$-$2405. 
The DE421 Solar System ephemeris of the Jet Propulsion Laboratory \citep{DE421} was 
used to transform the TOAs to the Solar System barycentre. 
Finally, the \textsc{tempo2} software package \citep{Hobbs2006} was used to fit a 
timing model to all TOAs, taking into account the astrometry, spin, and orbital motion of the pulsar.

\begin{figure}
\includegraphics[width=\linewidth]{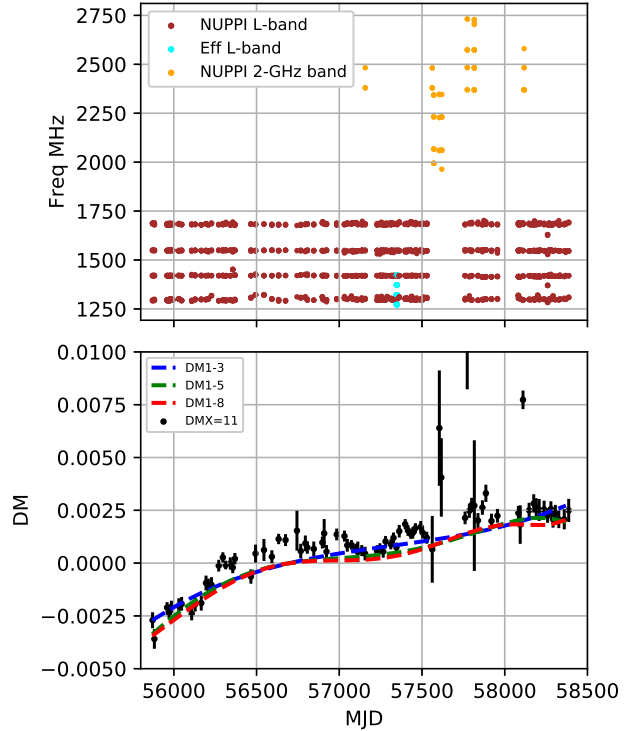}
\caption{Top panel: the temporal coverage of observing frequency of the coherently dedispersed
timing data used in this analysis. We create four sub-bands per observation when possible.  
The plot is colour-coded to show the different telescopes and backends, including
Nan\c{c}ay NUPPI at L-band (brown), Nan\c{c}ay NUPPI at 2\,GHz (orange) and 
Effelsberg PSRIX (cyan).
Bottom panel: DM variation as a function of time, measured by fitting for DMX with a window of 11\,d (black) in \textsc{tempo2}. There are small timescale DM variations that cannot be modeled properly using DM derivatives (blue: up to the third derivatives; green: up to the 5th derivative; red: up to the 8th derivative).}
\label{fig:DM}
\end{figure}

As mentioned in \citet{Ng2014}, PSR~J1811$-$2405 lies very close to the ecliptic 
plane with ecliptic coordinates $(\lambda, \beta) = (272^{\circ}.586, -0^{\circ}.675)$. 
With a $\lambda$ so close to 270$^{\circ}$, the uncertainty in $\beta$ will 
 only pertain to the declination when transforming in equatorial coordinates. 
We hence fixed proper motion in declination at zero for the rest of the analysis.
The parallax was also fixed at zero as the timing precision of our data is not high enough 
for measuring parallax with significance.

\begin{figure}
\includegraphics[width=3.2in]{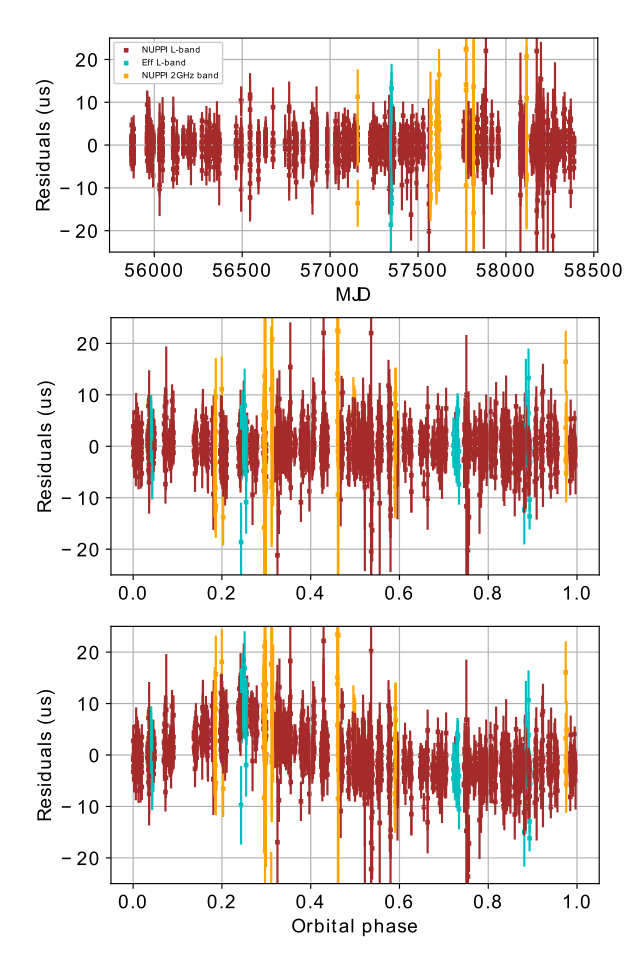}
\caption{Timing residuals for PSR~J1811$-$2405 as a function of MJD (top) and orbital phase (middle, bottom). TOAs are colour-coded to show different telescopes and backends, including 
Nan\c{c}ay NUPPI at L-band (brown), Nan\c{c}ay NUPPI at 2\,GHz (orange) and  
Effelsberg PSRIX (cyan).
20\,CM (blue), Parkes 50\,CM (cyan).
The middle panel shows the best-fit residuals when all parameters including the Shapiro delay are fitted for. 
The bottom panel shows the residuals if $\sin{i}$ and $m_{\rm{c}}$ are set to zero and all other 
parameters fixed at their best fit values.
The expected characteristic signature can be seen at orbital phase 0.25.
The error bars represent the 1-$\sigma$ uncertainties of the TOA measurements.}
\label{fig:residual}
\end{figure}

A careful modelling of the Dispersion Measure (DM) is very important for this work
because any unaccounted temporal DM variations can degrade the precision of a Shapiro delay measurement.
The fact that PSR~J1811$-$2405 is on the ecliptic plane also means it is more
susceptible to solar wind DM contribution. 
We used the solar wind model implemented by default in \textsc{tempo2} (where the density of
electrons is proportional to $1 / r_{\odot}^2$, with $r_{\odot}$ being the distance to the Sun,
see \citealt{tempo2}). But instead of the default value of 4 electrons/cm$^{3}$ at the
Earth's distance from the Sun proposed by \cite{tempo2}, we used 5~electrons/cm$^{3}$, which
was found to yield slightly better results by \citet{NANOGrav11}.
The timing baseline of PSR~J1811$-$2405 coincides in time with the NANOGrav
timing baseline, thus warranting the use of the same electron model.
We disregarded any observations that were taken with a solar elongation angle $<15^{\circ}$.

This simple model is not enough to describe DM variations from the
Solar wind, and it certainly cannot describe DM variations caused by the ionised interstellar 
medium. In order to do that, we must in addition use the
DMX parameterization in the \textsc{tempo2} software package . This a piecewise linear fit of temporal DM variations \citep[see][for a detailed definition]{NANOGrav9}. It
can be used to measure DM variations with a bin size of several days. 
In the bottom panel of Fig.~\ref{fig:DM}, we used a 11-d DMX window and showed that there is a 
temporal DM variation of the order of 0.005\,cm$^{-3}$\,pc over the course of our 
timing data. Alternatively, we also attempted to model this DM variation using DM derivatives. However, small timescale variations seen in Fig.~\ref{fig:DM} meant that even including up to the 8th DM derivative, the model still deviates visibly from the data. We hence concluded that DMX appears to be a more reasonable approximation of the DM trend compared to using DM derivatives.
We note that neither DM model has predictive power outside the range of data we have here.

\begin{table}
    \centering
  \caption{\textsc{tempo2} best-fit parameters for PSR~J1811$-$2405. Values in parentheses are the nominal 1-$\sigma$ uncertainties in the last digits.}
 \begin{minipage}{9cm}
\begin{tabular}{p{5.4cm}p{2.3cm}}
\hline
\multicolumn{2}{c}{Spin and astrometric parameters} \\
\hline
Right ascension, $\alpha$ (J2000) & 18:11:19.854050(19)\\
Declination, $\delta$ (J2000) & $-$24:05:18.422(10) \\
Proper motion in R.A., $\mu_{\alpha}$ (mas\,yr$^{-1}$) & 0.53(6) \\ 
Spin period, $P$ (ms) & 2.66059327687744(2) \\
Period derivative, $\dot{P}$ & 1.33756(3)$\times10^{-20}$ \\
Dispersion measure, DM (cm$^{-3}$\,pc) & 60.6196(2) \\
Rotation measure (rad\,m$^{-2}$) & 21(9) \\
\hline
\multicolumn{2}{c}{Binary parameters}\\
\hline
Orbital model & ELL1H \\
Orbital period, $P_{\rm{orb}}$ (days) & 6.27230196915(11)\\
Projected semi-major axis, $x$ (lt-s) & 5.7056569(8) \\
Epoch of ascending node, $T_{\rm{asc}}$ (MJD) & 56327.90581153(2) \\
$e \sin \omega$, $\epsilon_{1}$ & 9.9(9)$\times10^{-7}$\\
$e \cos \omega$, $\epsilon_{2}$ & 5.0(3)$\times10^{-7}$\\
Orthometric amplitude, $h_{3}$ ($\upmu$s) & 0.68(6) \\
Orthometric ratio, $\varsigma$ & 0.81(4) \\ 
\hline
\multicolumn{2}{c}{Derived parameters}\\
\hline
Inferred eccentricity, $e$ & 1.11(8)$\times10^{-6}$\\
Longitude of periastron, $\omega$ ($^{\circ}$) & 62(2) \\
Mass function, $f$ ($M_{\sun}$) & 0.00506927(2) \\
Companion mass, $m_{\rm{c}}$ ($M_{\sun}$) & 0.26(6) \\
$\sin{i}$ & 0.978(12) \\
Orbital inclination from Bayesian analysis, $i$ ($^{\circ}$) & $76.2\substack{+2.8 \\ -3.2}$\\
Companion mass from Bayesian analysis, $m_{\rm{c}}$ ($M_{\sun}$) & $0.31\substack{+0.08 \\ -0.06}$ \\
Pulsar mass from Bayesian analysis, $m_{\rm{p}}$ ($M_{\sun}$) & $2.0\substack{+0.8 \\ -0.5}$\\
DM-derived distance (kpc)$^{\dagger}$ & 1.8 \\
Intrinsic period derivative, $\dot{P}_{\rm{int}}$ & ${1.284(15)\times10^{-20}}^{*}$  \\
Characteristic age, $\tau_{\rm{r}}$ (Gyr) & 3.15 \\
Characteristic dipole surface magnetic field & 1.9 \\
strength at equator, $B_{\rm{eq}}$ ($10^{8}$ G)\\
Spin down power, $\dot{E}$ ($10^{34}$ erg s$^{-1}$) & 2.7 \\
\hline
\multicolumn{2}{c}{Timing model} \\
\hline
Binary model & ELL1 \\
Solar System ephemeris & DE421 \\
Timing epoch (MJD) & 56330 \\ 
First TOA (MJD) & 55871.6 \\ 
Last TOA (MJD) & 58386.9 \\ 
Weighted RMS residuals ($\upmu$s) & 1.81 \\
Reduced $\chi^{2}$  & 1.3 \\
\hline \label{tab:timing}
 \end{tabular}
\vspace{-0.5\skip\footins}
 \begin{flushleft}
$^{*}$ We have fixed the unconstrained proper motion in declination ($\mu_{\delta}$) at zero because PSR~J1811$-$2405 is very close to the ecliptic plane. The transverse velocity is therefore also not measurable. The derived $\dot{P}_{\rm{int}}$ represents an upper limit, obtained limit without correcting for any Shklovskii contribution \citep{Shk1970} in $\mu_{\delta}$.\\
$^\dagger$ Both the NE2001 \citep{NE2001} and the YMW2016 \citep{YMW2016} electron density models yield the same distance estimate.
\end{flushleft}
\end{minipage}
\end{table}

Since PSR~J1811$-$2405 has a very circular orbit with eccentricity of the order of $10^{-6}$, 
we used the ELL1 and ELL1H orbital models \citep{Lange2001,Freire2010} to characterize it.
In the more commonly used orbital models, like the BT \citep{BT} and DD \citep{DD} models,
there is a large covariance between two orbital parameters, the longitude of periastron $(\omega)$ and 
the epoch of periastron $\rm(T_{0})$; this becomes extremely large for 
small orbital eccentricities. The ELL1 and ELL1H models avoid this by
replacing those parameters and the orbital eccentricity $e$ by the Laplace-Lagrange parameters 
($\epsilon_{1}=e\sin\omega$ and $\epsilon_{2}=e\cos\omega$) and the time of ascending 
node passage $(\rm{T_{asc}})$, these have very small correlations among
themselves. The ELL1 and ELL1H models are approximate, they should only be used when the
ignored term (with amplitude $x.e^2\, = \, 7.0\, \times \, 10^{-12}$ lt-s)
is smaller than the timing precision ($T_{\rm rms} / \sqrt{N} \, = \, 3.6 \, \times \, 10^{-8} $ s);
this is certainly the case for PSR~J1811$-$2405. This implies that ignoring that term
will have, in this case, no other consequences, in particular it will have
no effect on the measurement of the Shapiro delay. 

Towards the end of the timing analysis when the reduced $\chi^{2}$ is close to one, 
we can assume that the timing model provides a reliable fit to the data.
As a last step, we compensated for any remaining systematics by calculating 
backend-specific weighing correction (also known as `EFAC' in 
\textsc{tempo2}). These coefficients were applied to scale the TOA uncertainties 
such that
the reduced $\rm\chi^2$ is unity for each individual data set.
This procedure yields more conservative and realistic estimates of the uncertainties in the timing parameters.

\begin{figure*}
\centering
\setlength\fboxsep{0pt}
\setlength\fboxrule{0pt}
\fbox{\includegraphics[width=16cm]{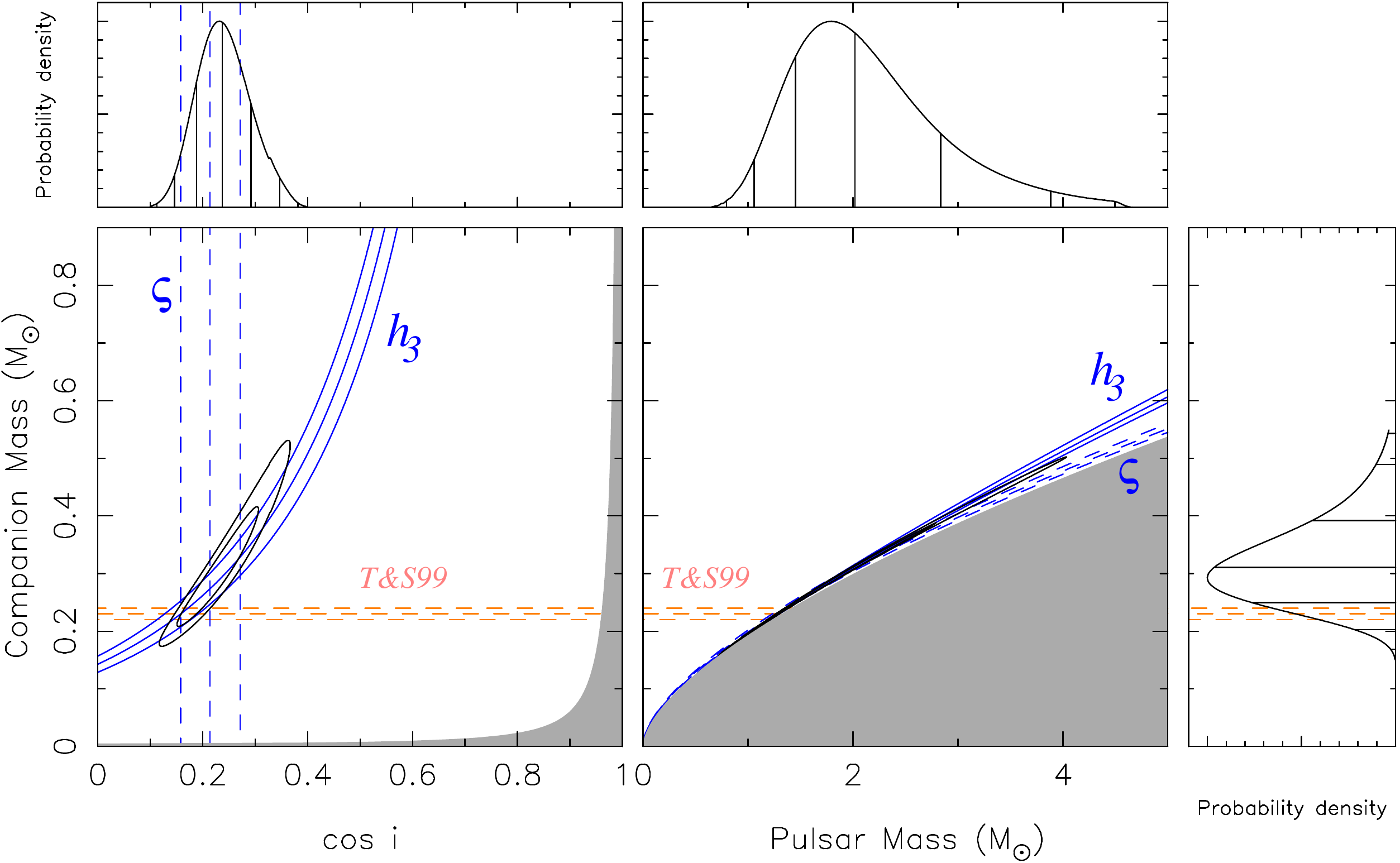}}
\caption{Shapiro delay constraints on the masses and orbital inclination of PSR~J1811$-$2405. 
The black contours contain 68.27 and 95.45 per cent of the total probabilities of the 
2-D pdfs from the Bayesian $\chi^{2}$ analysis.
The blue lines correspond to the nominal and $\pm1$-$\sigma$ uncertainties associated 
with the two orthometric parameters, $\varsigma$ (dashed line) and $h_{3}$ (solid line).
The pink lines labelled as T\&S99 represent the range of companion masses predicted by 
\citet{Tauris1999} for the orbital period of this system. 
Left: $\cos i-m_{\rm{c}}$ plane. The gray area is excluded by $m_{\rm{p}} > 0$. 
Right: $m_{\rm{p}}-m_{\rm{c}}$ plane. The gray area is excluded by $\sin{i} \le 1$.
The lateral panels are projected 1-D pdfs of $\cos{i}$, $m_{\rm{p}}$ and $m_{\rm{c}}$
respectively.}
\label{fig:mass}
\end{figure*}

The timing solution of PSR~J1811$-$2405, obtained using the ELL1H model, can be found in Table~\ref{tab:timing}. 
Only coherently dedispersed timing data from  Nan\c{c}ay and Effelsberg were employed in the timing analysis 
because they have the highest precision and represent the most homogeneous data set.

The main new result from the timing is a clear detection of the Shapiro delay. 
Fig.~\ref{fig:residual} compares the timing residuals when Shapiro delay is 
and is not taken into account. 
The signature sharp peak of a Shapiro delay can be seen clearly around phase 0.25, 
where the Earth-pulsar line of sight passes nearest to the companion 
as defined by the ELL1 and ELL1H binary models. 

In the ELL1 model
the Shapiro delay is characterized by two post-Keplerian parameters, the
range ($r$) and shape ($s$) \citep{Damour1992,Will1993}, where, assuming that General Relativity (GR)
is the correct description of gravity, $r=T_{\odot}m_{\rm{c}}$ and $s=\sin{i}$,  
(here $T_{\odot} \equiv G M_{\odot} c^{-3} = 4.9254909476412675\,\upmu$s is the 
mass of the Sun in units of time).
From these two parameters, we obtain $m_{\rm{c}}=0.26(6)\,M_{\odot}$ 
and $\sin{i}=0.978(12)$.
In the ELL1H model the Shapiro delay is described by two different PK parameters,
the orthometric ratio $\varsigma$ and amplitude $h_3$. Assuming GR, these are given by
$\varsigma=\frac{\sin{i}}{1+|\cos{i}|}$ and $h_{3}=T_{\odot}m_{\rm{c}}\varsigma^{3}$ \citep{Freire2010}.
These two parameters have much smaller correlation than $r$ and $s$, and hence provide a better description of the $m_{\rm{c}}$ and $i$ constraints 
derived from the Shapiro delay, particularly when the signal is weak. We report a highly significant 
(11-$\sigma$) $h_{3}$ of 6.8(6)$\times10^{-7}$  and a 16-$\sigma$ measurement of $\varsigma=0.81(5)$;
these yield mass and inclination estimates similar to those derived in the ELL1 model.
In Fig.~\ref{fig:mass}, we can see that these two parameters provide a good description of the regions
of the $\cos{i}-m_{\rm{c}}$ plane where the system is most likely to be located.
A detailed analysis of the mass and inclination constraints is presented in section \ref{sec:pmass}.

\section{RESULTS AND DISCUSSION} \label{sec:results}

\subsection{Polarization study} \label{sec:pol}
The Effelsberg timing observations recorded the four Stokes parameters in each frequency channel and 
thus can be used to study the polarization profile. 
We polarization calibrated each observation for the differential gain 
and phase between the feeds 
with an observation of the noise diode coupled to the receptors in the feeds. 
We made sure that the polarization calibration is taken adjacent to the targeted pulsar observations. 
In additional to the Effelsberg data sets, PSR~J1811$-$2405 was also followed-up at the Parkes 64-m radio telescope initially after its discovery. 
Incoherently-dedispersed Parkes Digital Filterbank systems (DFB) is available at L-band. 
A few observations have also been taken using the 
10/50\,cm receiver \citep{1050CM} 
which allow for the study of pulsar profile variations across frequencies.
The Parkes observations were flux calibrated by using an averaged observation of Hydra~A 
and the Effelsberg observation with the quasar~3C353. 

\begin{figure}
\includegraphics[width=3in]{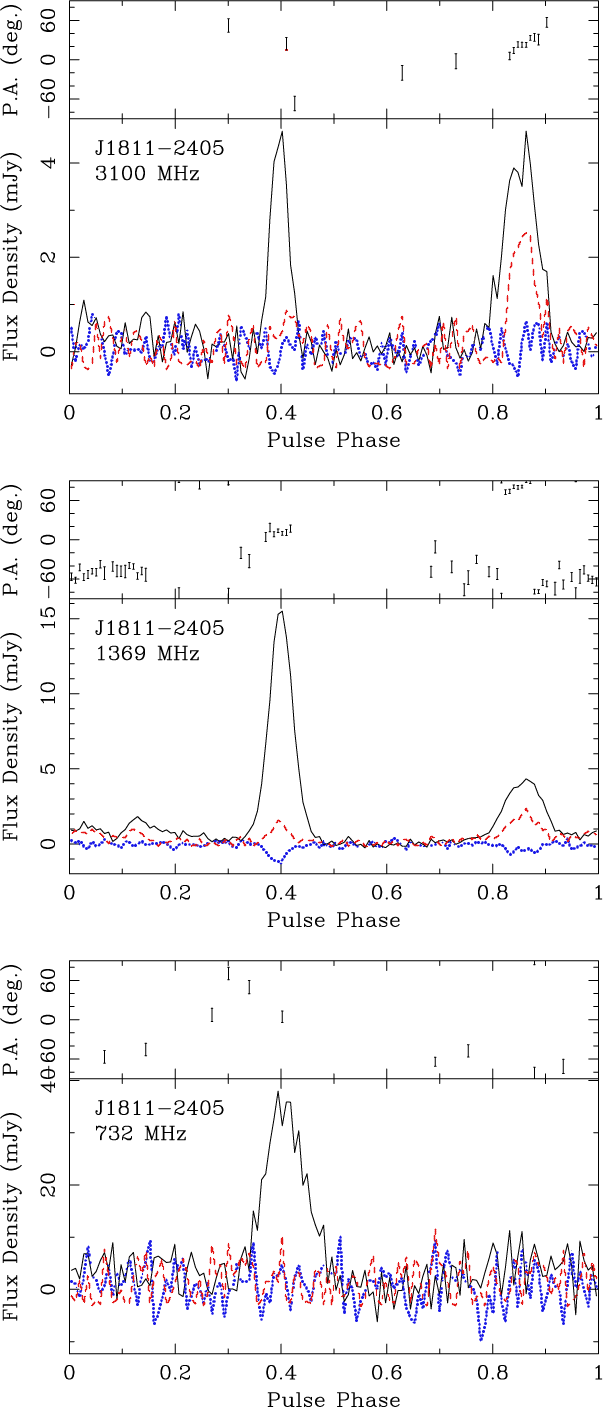}
\caption{Polarization profiles of PSR~J1811$-$2405 at (top) 3100\,MHz from Parkes, (middle) 1369\,MHz from Effelsberg
and (bottom) 732\,MHz from Parkes. 
The upper panel of each figure shows the RM-corrected PA variation in longitude with respect 
to the celestial north. Only PAs with signal-to-noise ratios $>3$ are shown.
The lower panel shows the integrated profile of total intensity (black solid line), linear 
polarization (red dashed line) and circular polarization (blue dotted line). }
\label{fig:pol}
\end{figure}

Fig.~\ref{fig:pol} shows the integrated polarization profiles of PSR~J1811$-$2405 
in total intensity, linear and circular polarization.
We have arbitrarily aligned the main pulse at phase 0.4 across the three 
available observing frequencies. 
Our best 732\,MHz profile comes from the co-adding of 4.1\,h of Parkes APSR coherently dedispersed data. 
Our best L-band profile comes from the co-adding of 14.6\,h of Parkes DFB incoherently dedispersed data.
At 3100\,MHz, a total of 4.7\,h of incoherently-dedispersed Parkes DFB observations are available.
We attempted to measure the observed Faraday rotation by fitting the position angle (PA) 
variations at L-band across the 256\,MHz band
and obtained a rotation measure (RM) of $21(9)$\,rad\,m$^{-2}$. The profiles shown in Fig.~\ref{fig:pol} have had their RM corrected.

PSR~J1811$-$2405 has a profile comprising two main components: a main pulse and an interpulse. 
The interpulse appears roughly 0.45 in phase after the main pulse.
We can also see that there is significant profile evolution across observing 
frequencies, as the strength of the interpulse decreases from being almost as bright as the 
main pulse at 3100\,MHz to being almost unidentifiable at 735\,MHz. 
The emission of the main pulse changes handedness in circular polarization which is 
most obvious at 1369\,MHz. 
The interpulse is strongly linearly polarized, especially at 3100\,MHz. 
PSR~J1811$-$2405 appears to have higher flux at lower observing frequencies, 
with a peak flux density of only a few mJy at 3100\,MHz but over 60\,mJy at 732\,MHz. 
Such a steep spectral index is typical of pulsars.

\begin{figure}
\includegraphics[width=3.25in]{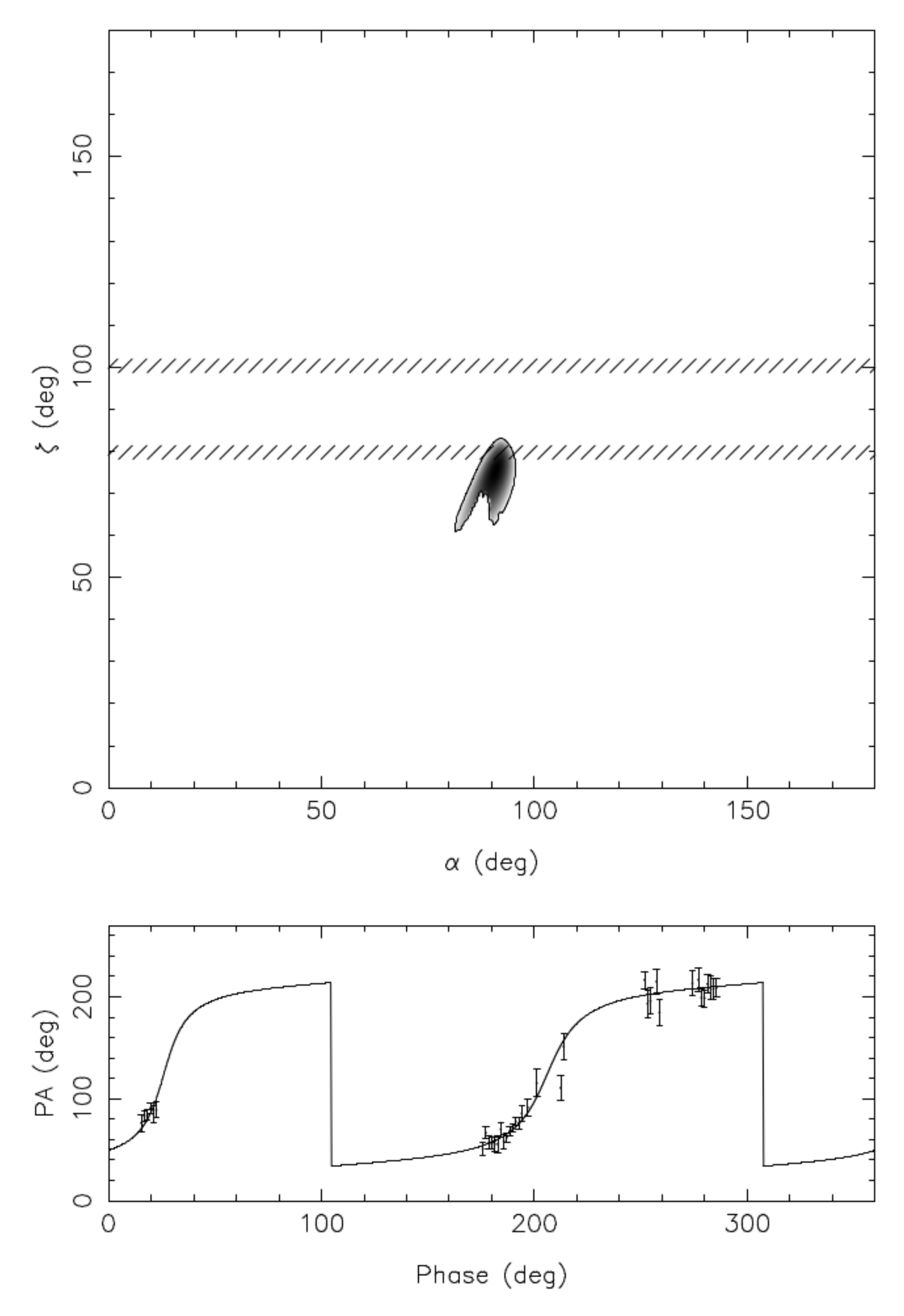}
\caption{Main panel: System geometry for PSR~J1811$-$2405 from a least-squares fit of the RVM
to the PA. The contour indicates 1-$\sigma$ best-fit region of $\alpha$ and $\zeta$. The 
orbital constraints from our Shapiro delay measurement are marked as two horizontal bands. 
Bottom panel: The corresponding RVM fit for 
$\alpha=92^{\circ}$ and $\zeta=79.8^{\circ}$ 
is shown. 
An orthogonal shift is applied for the 
second group of points, as well as a $180^{\circ}$ phase shift between the first and third group.}
\label{fig:rvm2}
\end{figure}

We modeled the observed PA swing according to the Rotating Vector Model \citep[RVM;][]{RVM1969}. 
As detailed in \citet{Handbook}, the RVM can in principle provide insights on the 
magnetic inclination angle, $\alpha$, as well as the viewing angle between the 
observer and the spin axis, $\zeta$. 
We followed the procedure of a least-squares fit similar to that described in \citet{Berezina}, by 
stepping through a range of $\alpha$ and $\zeta$ while simultaneously minimizing the reference
phase $\Phi_{0}$ and the reference position angle $\Psi_{0}$ of the 
RVM at each grid point. This results in the 1-$\sigma$ $\chi^{2}$ contour shown in Fig.~\ref{fig:rvm2}.
From the system parameters, we can expect PSR~J1811$-$2405 to have undergone an extended recycling process, which not only transferred masses but also orbital angular momentum.
As a result of this accretion phase, the spin angular momentum of the pulsar aligned with the orbital angular momentum over time. Hence, in order for the pulsar beam to be visible to a
terrestrial observer, the viewing angle $\zeta$ must be roughly consistent with the orbital 
inclination angle \citep[within the uncertainty of the angular radius of the pulsar beam, $\rho$, c.f.][]{Guillemot2014}.
We can thus expect $\zeta \approx i$. From the Shapiro
delay measurements, we can only determine $\sin i$, hence both $i$ and $180^{\circ}-i$ are possible. With this in mind, we mark the value of $i$ and $180^{\circ}-i$ from our pulsar timing Shapiro delay on Fig.~\ref{fig:rvm2} as two horizontal bands, at 79$^{\circ}$ and 101$^{\circ}$ respectively. The width of these horizontal bands represents the uncertainty on $i$.

The intersection region of the polarimetry and orbital inclination favours a solution with 
$\alpha=92^{\circ}$ and $\zeta=79.8^{\circ}$, 
indicating that the true underlying orbital inclination angle is 
$i = 180^{\circ}-79.8^{\circ} \sim 100^{\circ}$.
This $\alpha$ value also indicates that PSR~J1811$-$2405 is an orthogonal rotator, 
which is consistent with the fact that we see an interpulse from the pulse profile. 
It is sometimes possible to obtain further constraints on $\alpha$ through a fitting of the pulse width, together with an assumption of a filled emission beam. However, the duty cycle of PSR~J1811$-$2405 is relatively low and only the width of the main peak can be measured, which limits the reliability of this analysis. This is not unusual for MSPs, especially when low-level components are present as in the case of PSR~J1811$-$2405.

\citet{Guillemot2014} analysed a sample of MSPs with $\zeta$ constraints, and found marginal evidence for different viewing angle distributions between gamma-ray-detected and undetected energy and nearby MSPs. They postulated that gamma-ray-undetected MSPs are seen under small viewing angles. PSR~J1811$-$2405 is detected in gamma rays and is seen under a large viewing angle, and thus seems to follow the trend.
We also compared the gamma-ray profile of PSR~J1811$-$2405 with the light curve 
models presented in \citet{Romani2010}. 
Qualitatively, we found good agreement with their light curve corresponding to 
($\alpha,\zeta)=(90^{\circ},80^{\circ})$. We do not see obvious preference 
between the `two pole caustic' (TPC) and the outer gap (OG) model.

\subsection{Pulsar mass constraint from the Shapiro delay measurements} \label{sec:pmass}
As described by \citet{Splaver2002}, 
we can estimate the masses and their uncertainties for the pulsar ($m_{\rm{p}}$) 
and its companion 
by performing a Bayesian $\chi^{2}$ analysis in the $m_{\rm{c}} - \cos{i}$ plane.
We stepped through a regular grid ranging from $ 0.15\,M_{\odot} \le m_{\rm{c}} < 0.55\,M_{\odot} $ and from 
$0.1 \le \cos{i} < 0.4$, involving a total of 150$\times$400 trial values. 
The assumption of a flat $\cos{i}$ plane can be justified if we consider 
the binary orbit to be randomly oriented in space.
For each of these $m_{\rm{c}}-\cos{i}$ pairs, 
we calculated the corresponding $r$ and $s$ to be held fixed. We fit for all 
other timing parameters and record the final $\chi^{2}$.
Fig.~\ref{fig:mass} shows the resultant $\chi^{2}$ map, from which 2-D 
probability distribution functions (pdfs) can be derived.
The black contours represent the 68.27 and 95.45 per cent of the total probability, 
which is consistent with that from the fitting of $h_{3}$ and $\varsigma$ 
(blue lines) using the orthometric parametrization.

We can then marginalize the 2-D pdfs by projecting them 
onto the $m_{\rm{c}}$, $\cos{i}$ and $m_{\rm{p}}$ axes.
From these 1-D pdfs, we obtain  
$m_{\rm{p}}=2.0\substack{+0.8 \\ -0.5}\,M_{\odot}$,
$m_{\rm{c}}=0.31\substack{+0.08 \\ -0.06}\,M_{\odot}$, 
and $i=76^{\circ}.2\substack{+2^{\circ}.8 \\ -3^{\circ}.2}$\,,
where the quoted values are the medians together with the $\pm1-\sigma$ interval.
The current value of $m_{\rm{p}}$ is not yet very well constrained 
but could be improved with some more years of pulsar timing.

The detection of other post-Keplerian parameters such as the orbital 
period variation ($\dot{P}_{\rm{orb}}$) and
the variation in the longitude of periastron ($\dot{\omega}$) could, in a more
compact and eccentric system, have 
helped to narrow down the pulsar mass range and to conduct
self-consistency tests of GR and other theories of gravity. 
For this system, 
a $\dot{P}_{\rm{orb}}$ of the order of $10^{-16}$ is predicted from 
orbital decay due to GR, which is too 
small to be easily separated from kinematic effects due to the Galactic acceleration and 
the Shklovskii effect \citep{Shk1970}, hence we do not expect to measure any intrinsic $\dot{P}_{\rm{orb}}$ in the near future. 
If we assume a pulsar mass of 1.5\,$M_\odot$, for an orbital inclination of $79^{\circ}$, one can expect an $\dot{\omega}$ of 0.0136$^{\circ}$\,yr$^{-1}$.
Currently, we measure $\omega$ with a precision of 0.7$^{\circ}$. This means we will have to wait at least 50\,yr to
achieve a 1-$\sigma$ measurement of $\dot{\omega}$.

A more promising improvement can be achieved with the MeerKAT \citep{Bailes2018} and even better, when the Square Kilometre Array \citep[SKA;][]{Stappers2018} comes online, 
which will most certainly  provide much better constraints on the Shapiro delay measurement.

\subsection{Evolution model}
\citet{Tauris1999} conducted numerical calculations on the non-conservative evolution 
of close binary systems. They suggested that for diverging LMXBs with a donor mass $<2\,M_{\odot}$
and a $1.3\,M_{\odot}$ accreting neutron star, if the orbital period is greater than 
the orbital bifurcation period, i.e. $P_{\rm{orb}} > P_{\rm{bif}}$ ($\simeq$2\,d), then 
the system should follow a positive correlation between the orbital period and the mass 
of the final He-WD companion, quantified by Equation~(20) and (21) in \citet{Tauris1999}.

PSR~J1811$-$2405 belongs to this type of diverging system, with $P_{\rm{orb}}=6.27$\,d. 
In Fig.~\ref{fig:mass}, we overplot the range of companion masses predicted by 
\citet{Tauris1999} for the orbital period of this system.
The measurement of $m_{\rm{c}}$ is consistent with the prediction of \citet{Tauris1999}, but because of its large uncertainties, it cannot yet test the relation. We note that if the companion has the predicted mass, then the Shapiro delay parameters would imply a pulsar mass around $1.3\, M_{\odot}$.
It would also mean a $\cos{i}$ of $\pm$0.16, which equals to $i\sim$81$^\circ$ or 99$^\circ$. This is in  agreement with the polarization estimate, and matches the orthogonal rotator scenario.

\section{Summary} \label{sec:conclusion}
We present an updated radio timing solution for the binary system PSR~J1811$-$2405. 
An extended timing campaign and high precision coherently dedispersed observations have allowed for the first detection of the relativistic Shapiro delay.
We measured the orthometric amplitude $h_{3}$ and ratio $\varsigma$ 
with high significance.
By conducting a Bayesian $\chi^{2}$ analysis, we obtained  
constraints on the companion mass to be 
$m_{\rm{c}}=0.31\substack{+0.08 \\ -0.06}\,M_{\odot}$
and a less well constrained pulsar mass of 
$m_{\rm{p}}=2.0\substack{+0.8 \\ -0.5}\,M_{\odot}$.

The companion mass is in agreement with the theoretical $m_{\rm{c}}-P_{\rm{orb}}$ correlation
obtained by \citet{Tauris1999}. 
From the polarization study, we obtained a solution of $\alpha=92^{\circ}$ and 
$\zeta=79.8^{\circ}$, indicating that the true underlying orbital inclination 
angle is $i\sim100^{\circ}$, consistent with PSR~J1811$-$2405 being an orthogonal rotator.
The high timing precision and sharp profile of PSR~J1811$-$2405 make it a good candidate 
to be included in a pulsar timing array to aid the gravitational wave detection effort.
With a longer timing baseline we can then expect improvements in the precision of the 
pulsar mass measurement. 

\section{acknowledgements}
The Parkes Observatory is part of the Australia Telescope National Facility, 
which is funded by the Commonwealth of Australia for operation as a National Facility managed by CSIRO. 
This work was partly based on observations with the 100-m telescope of the MPIfR 
(Max-Planck-Institut f\"{u}r Radioastronomie) at Effelsberg. 
The Nan\c{c}ay radio observatory is operated by the Paris Observatory, associated with the 
French Centre National de la Recherche Scientifique (CNRS). 
We also thank Marina Berezina, Eleni Graikou and Gregory Desvignes for advice on Effelsberg data calibration, as well as Andrew Cameron for carefully reading the manuscript.

\bibliographystyle{mn2e}

\label{lastpage}
\end{document}